\documentclass[aps,prstab,reprint,groupedaddress]{revtex4-1}
\usepackage{amssymb, amsmath}
\usepackage{graphicx}
\usepackage{color}
\usepackage{xspace}
\usepackage{titlesec}

\newcommand{\degree}{\ensuremath{^\circ}\xspace}
\newcommand{\htp}{\ensuremath{\mathrm{H}_2^+}\xspace}
\newcommand{\BE}[0]{\begin{equation}}
\newcommand{\EE}[0]{\end{equation}}
\newcommand{\BEA}[0]{\begin{eqnarray}}
\newcommand{\EEA}[0]{\end{eqnarray}}

\newcommand{\nuebar}{\ensuremath{\bar{\nu}_e}\xspace}

\mathchardef\mhyphen="2D

\newcommand{\opal}{\textsc{OPAL}\xspace}

\newcommand{\DD}{DAE$\delta$ALUS\xspace}

\titleformat{\section}
{\center\normalsize\bfseries\uppercase}{\thesection}{1em}{}

\titlespacing\section{0pt}{12pt plus 4pt minus 2pt}{6pt plus 4pt minus 2pt}
\titlespacing\subsection{0pt}{12pt plus 4pt minus 2pt}{6pt plus 4pt minus 2pt}
\titlespacing\subsubsection{0pt}{12pt plus 4pt minus 2pt}{6pt plus 4pt minus 2pt}

\begin{document}
\title{A Response to arXiv:1512.09181, ``Space Charge Limits in the \DD DIC Compact Cyclotron''}
\author{Janet Conrad}
\email{Corresponding author: conrad@mit.edu}
\author{Mike Shaevitz}
\author{Andreas Adelmann}
\author{Jose Alonso}
\author{Luciano Calabretta}
\author{Daniel Winklehner}
\affiliation{for the \DD Collaboration}
\date{\today}

\begin{abstract}
This document addresses concerns raised about possible limits, due to 
space charge, to the maximum \htp ion beam current that can be injected 
into and accepted by a compact cyclotron. 
The discussion of the compact cyclotron is primarily within the context
of the proposed \DD and IsoDAR neutrino experiments.
These concerns are examined by the collaboration and addressed 
individually.\
While some of the concerns are valid, and present serious challenges
to the proposed program, the collaboration sees no immediate showstoppers.\
However, some of the issues raised clearly need to be addressed carefully -
analytically, through simulation, and through experiments. 
In this report, the matter is discussed, references are given to work 
already done and future plans are outlined.
\end{abstract}

\pacs{}

\maketitle

\section{Introduction}
\DD \cite{alonso:daedalus, aberle:daedalus, abs:daedalus,
calabretta:daedalus, conrad:daedalus} is a proposed experiment to measure CP violation in the neutrino sector.\ To provide the necessary
neutrino flux to obtain results in a 5 year measurement period, 10 mA of
protons need to be accelerated to 800 MeV/amu and impinge on the 
target where they create neutrinos through pion and muon decay-at-rest.\ 
The system is envisioned to be a chain of two cyclotrons: the compact 
60 MeV/amu \DD Injector Cyclotron (DIC), that acts as a pre-accelerator,
and the 800 MeV/amu \DD Superconducting Ring Cyclotron (DSRC).\
Using only the DIC, an isotope decay-at-rest experiment can be setup, producing a very
pure \nuebar beam and a decisive search for sterile
neutrinos can be anticipated.\ This project is called IsoDAR \cite{bungau:isodar,
adelmann:isodar, abs:isodar}.\
Among the challenges for the DIC are the strong space charge effects of 
such a high intensity beam.\ Space charge matters most in the Low 
Energy Beam Transport Line (LEBT) and during injection into the cyclotron.
Our concept to mitigate this problem is based on accelerating 5 mA of \htp
instead of 10 mA of protons, leading to the same number of nucleons on
target at half the electrical current as the remaining electron bound in 
the \htp molecular ion reduces the electrical current in the beam.
In addition, stripping of this electron leads to very clean extraction 
from the DSRC.


Recently, a note appeared on arXiv \cite{bm}, discussing the issue of 
space charge during injection into the DIC and estimating the maximum 
achievable \htp current in a compact cyclotron at about $200\: \mu$A.
This is far from the 5 mA of \htp beam current needed for the 
proposed experiment. The collaboration has 
carefully examined the presented arguments and found that, while there
certainly is merit to them, matters are over-simplified in the 
published note and moreover, it does not take into account important 
published results \cite{Yang201384, winklehner:bcs_tests}.

In Section \ref{sec:discussion} the arguments are discussed in detail.

\section{Executive Summary}
In this note we carefully consider the arguments presented in
\cite{bm}, and find that while some of the points are well-taken, the 
foundations are too weak for the strong statement that the maximum
\htp current to be expected from our system will never exceed 
$200\: \mu$A.\

We agree that the perveance argument must be used with caution, 
and that comparing our \DD Injector Cyclotron (DIC) with a high-current 
H$^-$ cyclotron is not appropriate.
Space charge forces are directly related to the bunch charge density, 
and in  H$^-$ machines, with substantially larger phase acceptance,
the longitudinal extent of the bunch is considerably greater. 
In contrast, such a large phase acceptance cannot be tolerated in the 
DIC in which a good turn separation is required at extraction.

We also agree that the critical elements of our concept are the injection 
and first few turns.
In our paper \cite{Yang201384} (not cited in \cite{bm}), we developed a 
beam dynamics model, from 1.5 MeV/amu to extraction at 60 MeV/amu, 
showing that the beam can be accelerated and extracted with acceptable 
beam loss on the extraction septum.\ We also show that the vortex motion 
of the bunch provides good longitudinal and radial focusing, resulting in 
a quasi stationary bunch, similar to the PSI Injector 2.
In a current research project, we are studying the beam dynamics through the 
axial injection channel, the spiral inflector, and acceleration through the first few turns to match, with the input conditions (at 1.5 MeV/amu) of the model developed in \cite{Yang201384}.

We have already demonstrated that a larger inflector can be built; 
we have, in fact, tested a 15 mm gap inflector with good 
transmission of \htp beams \cite{winklehner:bcs_tests} 
(not cited in \cite{bm}).
This refutes the argument made in \cite{bm} that 
``a bunch of 12 mm full 4$\sigma$ size would not fit through a 
reasonably-dimensioned inflector''.

Furthermore, \cite{bm} does not acknowledge that our higher injection energy, leading to larger radius of the first turn, and the use of four high power RF cavities will provide very rapid acceleration of the injected beam with additional vertical focusing. We anticipate that these elements  will be essential for the challenging matching process at the takeoff point of our published model in \cite{Yang201384}.

This said, it is clear that careful and precise PIC simulations are required to establish the feasibility of the necessary injection and matching conditions.\ We are embarking on a program to do exactly this.\ The tools are in place now, following the enhancement of \opal to include the 3D inflector 
geometry, and anticipate having suitable answers in an appropriate time frame.

\section{Discussion of the Concerns}
\label{sec:discussion}
In discussing the raised concerns, we will largely follow the structure
of \cite{bm} and add additional context as needed.

\subsection{Generalized Perveance}
In our publications, we often use the generalized perveance $K$ as a first order
estimate of the strength of the space-charge effects in various parts of
the \DD system (LEBT, DIC, DSRC). In \cite{reiser:beams}, $K$ is given by:
\begin{equation} 
K = \frac{q}{2 \pi \epsilon_0 m} \frac{\mathrm{I} (1-\gamma^2\mathrm{f}_\mathrm{e})}{\gamma^3 \beta^3},
\label{eqn:space-charge}
\end{equation}
with q, I, m, $\gamma$, and $\beta$ the 
charge, current, rest mass, and relativistic parameters of the particle beam, 
respectively, and $\mathrm{f}_\mathrm{e}$ the space charge compensation factor.\

In \cite[Section 1]{bm}, the applicability of the 
perveance argument is examined and it is pointed out that:
\begin{enumerate}
\item The comparison between the average beam currents of a compact 
\htp cyclotron and a compact H$^-$ cyclotron is erroneous, because space charge
is a local effect and the phase acceptance of an H$^-$ machine is much larger,
thus the bunches are longer and the peak currents are significantly lower.
\item We are injecting at roughly double the energy of typical compact 
cyclotrons (70 keV instead of 30 keV) which indeed leads to lower perveance, but
has the disadvantage that the transverse focusing is one quarter as large due
to the injection radius being twice as large.\
\end{enumerate}

\begin{figure*}
    \centering
    \includegraphics[width=1.0\textwidth]{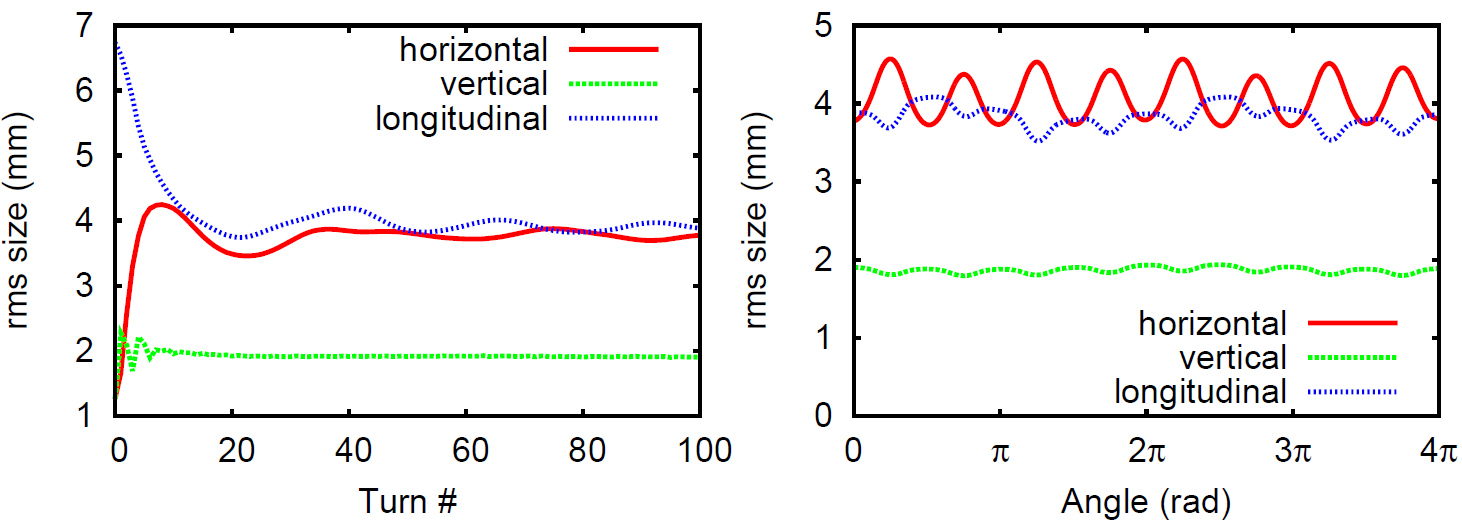}
    \caption{The rms size snapshot at 0\degree azimuth in 100 turns (left), 
             and the rms envelope in the last two turns (right). 
             From \protect\cite{Yang201384}.}
    \label{yang_fig2}
\end{figure*} 


With respect to item 1 above, we acknowledge that the comparison of H$^-$ 
cyclotrons with a large acceptance (TR-30 \cite{kuo:tr30} and CYCLONE-30 
\cite{vanderlinden:cyclone30}) and the 
DIC was made without taking the phase acceptance into account. 
\cite{bm} correctly states that the phase acceptance of 
H$^-$ cyclotrons is about 60-70\degree and in the DIC it is only 
20\degree. Thus the average current is compressed into a smaller bunch 
and our peak current is about 90 mA for 5 mA DC 
equivalent average current (5 mA $\cdot$ 360\degree / 20\degree).
In a H$^-$ machine the peak current would only be 30 mA .

With respect to item 2, we assume this argument is based on the
2D Model presented in \cite[Section 2.1]{bm} (briefly discussed 
in the next section here) and the envelope equation therein 
(cf. Equation \ref{eqn:envelope}).
Indeed, if one were to compare H$^-$ and \htp at the same velocity
and in essentially the same machine (same tunes, same acceleration, etc.)
and at the same current, according to the simple model,
the situation for \htp would be a factor 2 worse,
due to the decrease of the focusing strength with turn radius
squared. However, this argument does not truly apply to
the DIC for two reasons:
the design is indeed different and the envelope equation does not include
important effects of beam dynamics (see discussions in the following 
sections).


Clearly, previous comparison with existing
compact H$^-$ cyclotrons is obsolete and we should instead focus
on the design choices we made to accommodate the higher rigidity 
of \htp.

Ultimately, the feasibility of the project needs to be determined through 
careful analysis, a rigorous simulation study of the actual DIC design,
and experiments.

\subsection{A 2D Model?}
In \cite[Section 2.1]{bm} the envelope equation in terms of 
perveance and tune is investigated briefly. Cited directly from 
\cite{bm}:
\BE
x^{\prime\prime} = - \frac{\nu_x^2}{R^2}x + \frac{\epsilon_x^2}{x^3} + 
\frac{2K}{x+y}
\label{eqn:envelope}
\EE
with $x$, $y$ the radial and vertical envelope radii, $\nu_x$ the radial tune
and $R$ the orbit radius.

We agree with the statement in the last paragraph of this section, i.e.\
a continuous beam modeled with the envelope equation is a not suitable model.\
This is due to the fact that complex particle dynamics present in space charge dominated beams, leading to vortex-motion and the formation of a round distribution
in radial-longitudinal space, are not considered.

In addition, the DIC will differ from existing machines in several key 
aspects.
The spiral inflector and cyclotron vertical gap will be sufficiently 
large to let a larger beam pass through. 
The energy gain per turn will be 2.4 to 2.6 times 
higher compared to that in proton cyclotrons, to expedite exit from 
of the central region and to achieve earlier development of a 
stabilizing vortex motion, leading to a stationary distribution.
The DIC will have 4 double-gap dees and thus better vertical focusing
with $\nu_z = 0.5$ already at the 8th turn.

We thus conclude that all other statements made in Section 2 have no merit
and only a fully 3D treatment of the particle dynamics and experimental 
results (and the comparison of the two) can determine the maximum achievable 
current in the DIC.

\subsection{3D Modeling}
Knowing that injection will be a challenging task, we first investigated the space charge dominated beam transport starting from 1.5 MeV/amu up to extraction of the DIC using the particle-in-cell (PIC) code \opal \cite{adelmann:opal}.
These encouraging results are published in \cite{Yang201384} and 
were not referenced in \cite{bm}.\

In \cite{Yang201384} we deliberately started with a mismatched and collimated beam, to mimic mismatch and study the reordering of the
phase space (see Figure \ref{yang_fig2}).\ Indeed we could show clearly that an almost stationary distribution is formed and the beam can be extracted with tolerable losses.\ Inspection of \cite[Figure 6]{Yang201384} provides evidence that the turn separation at injection and extraction is sufficient. 

To conclude, we have carefully investigated the matching process that leads to a stationary and extractable distribution in the DIC. However, we did not yet include the spiral inflector in the model, this is part of an ongoing research project. The
model presented in \cite{bm}, based on second order moments of the charge distribution, will not be able to reproduce in detail 
the complicated space charge dynamics, mostly because of the lack of non-linearities. Here we stress the fact that only a full 3D
particle model can show us the final limits. 

\subsection{A Possible RFQ Injection Scheme}
In  \cite[Section 3.1]{bm}, the author briefly discusses our 
ongoing investigation regarding the use 
of an RFQ as an injector to the DIC.\ The general idea of an RFQ
direct injection scheme for compact cyclotrons was first published 
in 1981 by R.W. Hamm \cite{hamm:rfq1} and recently investigated 
together with Hamm for use in the DIC \cite{winklehner:rfq1}.
This recent investigation was prompted by the fact that the ion source used
in the \htp injection tests at Best Cyclotron Systems, Inc.\
\cite{winklehner:bcs_tests} did not deliver the necessary current. 
As is pointed out in \cite{abs:isodar}, our primary design 
uses a conventional LEBT with a new improved ion source.
The RFQ is an alternative design under investigation.

As we mention in \cite{winklehner:rfq1}, the preliminary design produces a
beam that de-bunches longitudinally and grows rapidly in the 
transverse direction. We are currently investigating the matching of 
the RFQ output beam to the spiral 
inflector and the cyclotron using \opal.\ It was recognized
in \cite{winklehner:rfq1}
(and now confirmed in \cite{bm}) that additional
focusing and re-bunching will be necessary to make this system work.\
This is highly experimental and preliminary. However,
dismissing the idea out-of-hand would be a mistake, because if 
it works, it could drastically relax the requirements of the injector
ion source.

Regarding space charge compensation, it is true that transport through a
conventional LEBT would yield higher compensation
thereby leading to smaller emittances compared to the RFQ.\
However, starting at the entrance of the spiral inflector no compensation 
will be possible in either case due to the strong electric fields.\
Whether the increased emittance of the RFQ is prohibitively large will have to
be determined through careful simulations as well.

\subsection{Matching}
In \cite[Section 3.2]{bm}, it is stated that there is no way of matching the
inflector output to the first turn optics. 

Referring to the discussions and citations given above, we do not 
acknowledge the validity of the heuristic factor of 2 assumed in 
\cite{bm} without justification.
The ongoing effort to build a 3D beam dynamics model, 
including the spiral inflector and the full central region will 
allow us to understand the complicated matching process. 
In turn, this will provide a sound answer regarding the feasibility 
of our approach. 

\subsection{Theory Conclusions}
The concluding theory section of \cite[Section 5]{bm} claims 
that 150 pC (5 mA at 33 MHz) bunches are impossible for 2 reasons:
\begin{enumerate}
\item A bunch of 12 mm full 4$\sigma$ size would not fit through a `reasonably-dimensioned' inflector.
\item The resulting bunch occupies 54\degree RF.
\end{enumerate}
However, in \cite{winklehner:bcs_tests}
(not cited in \cite{bm}),
we had described a test cyclotron for 
\htp injection and the spiral inflector which has a gap of 15 mm. Tests showed
that we could inject $\approx$ 6 mA \htp DC, limited not by the cyclotron
central region, but by the ion source current and LEBT. 
These values corresponded very well to simulations using \opal. 
The simulations were later increased to 50 mA with similarly 
encouraging results.
In \cite{Yang201384} we showed that even an initially mismatched beam that occupies more than the 10\degree RF will undergo vortex motion and within
a few turns form a round distribution that can fit within this phase window.

The lessons we learned here are the following: a) the estimates presented in  \cite{bm} are very pessimistic and are (partially) refuted by the
experiment. b) From this fact we can also conclude that our 3D model is closer to the nature than simplified estimates.

\subsection{Envelope Evidence using {\tt TRANSOPTR}}
In \cite[Section 4]{bm} {\tt TRANSOPTR}, a code based on the 3D envelope equation, including linear space charge was used.
Several RFQ related scenarios were studied.\
These studies are certainly very interesting, however, the assumption of a fixed vertical tune of $\nu_{z} = 0.3$ and no acceleration does not represent the true nature of this complicated part of the DIC.\

Indeed the DIC is designed to work with 4 RF cavities, and the cavities can be operated, already from the first turn, with voltages of 70-80 kV, i.e.\ 16-30\% higher than the usual 60 kV used in the H$^-$ compact cyclotron.\ The use of higher voltage is just a consequence of the larger radius of the \htp cyclotron.\
Moreover, the larger gap mitigates the problem of a possibly 
large vertical size due to the initial $\nu_{z} = 0.3$. 
Additionally, the value of $\nu_{z}$  increases to 0.4 at the 4th turn 
and 0.5 at the 8th turn. 

\subsection{Ideally Placed Bunches with {\tt TRANSOPTR}}
In \cite[Section 5]{bm} the author uses {\tt TRANSOPTR} to 
investigate the evolution of ``ideally placed'' bunches.
In the final paragraph he concludes that the smallest bunch 
capable of containing the 150 pF would have to be roughly 6 mm
in radius. 
Here we refer to \cite{Yang201384}, not referenced in \cite{bm},
in which we did not start at turn one, but at 1.5 MeV/amu. 
In the near future we can also envision simulations
starting at turn 1, especially to study the possible 
collimation scenarios.

\section{Conclusions}
The limits presented in \cite{bm} are not the upper limits 
of the performance of compact cyclotron-based accelerator systems.
In this note we give evidence partially through published models and 
partially through experiments. 
Extending these limits is the goal of our research program.

We note that the DAEdALUS situation may be considered analogous to 
the early days of the Paul Scherrer Institut 590 MeV meson factory, 
then called SIN (Swiss Institute for Nuclear-research). Skeptics 
were quite vocal about this cyclotron system never exceeding currents 
of 100 $\mu$A \cite{johopc}. As is now history, through careful 
and methodical improvements and upgrades, the proton current is 
now over 2.4 mA, nearly a factor of 25 higher with respect 
to the pessimistic estimates of experts in the past.

\bibliography{Bibliography}

\end{document}